\documentclass[aps,prl,reprint]{revtex4-1}
\usepackage[ansinew]{inputenc}
\usepackage{graphicx}
\usepackage{textcomp}
\usepackage{hyperref}

\begin{document}

\title{An Elementary Quantum Network of Single Atoms in Optical Cavities}

\author{Stephan Ritter}
\email[To whom correspondence should be addressed. E-mail: ]{stephan.ritter@mpq.mpg.de}
\author{Christian Nölleke}
\author{Carolin Hahn}
\author{Andreas Reiserer}
\author{Andreas Neuzner}
\author{Manuel Uphoff}
\author{Martin Mücke}
\author{Eden Figueroa}
\author{Jörg Bochmann}
\altaffiliation[Present address: ]
{Department of Physics, University of California, Santa Barbara, CA 93106-9530, USA and California NanoSystems Institute, University of California, Santa Barbara, CA 93106-9530, USA}
\author{Gerhard Rempe}
\affiliation{Max-Planck-Institut für Quantenoptik, Hans-Kopfermann-Strasse 1, 85748 Garching, Germany}

\begin{abstract}
Quantum networks are distributed quantum many-body systems with tailored topology and controlled information exchange. They are the backbone of distributed quantum computing architectures and quantum communication. Here we present a prototype of such a quantum network based on single atoms embedded in optical cavities. We show that atom-cavity systems form universal nodes capable of sending, receiving, storing and releasing photonic quantum information. Quantum connectivity between nodes is achieved in the conceptually most fundamental way: by the coherent exchange of a single photon. We demonstrate the faithful transfer of an atomic quantum state and the creation of entanglement between two identical nodes in independent laboratories. The created nonlocal state is manipulated by local qubit rotation. This efficient cavity-based approach to quantum networking is particularly promising as it offers a clear perspective for scalability, thus paving the way towards large-scale quantum networks and their applications.
\end{abstract}

\maketitle

Connecting individual quantum systems via quantum channels creates a quantum network with properties profoundly different from any classical network. First, the accessible state space increases exponentially with the number of constituents. Second, the distribution of quantum states across the whole network leads to nonlocal correlations. Further, the quantum channels mediate long-range or even infinite-range interactions which can be switched on and off at will. This makes quantum networks tailor-made quantum many-body systems with adjustable degrees of connectivity and arbitrary topologies, and thus powerful quantum simulators. Open questions like the scaling behaviour, percolation of entanglement \cite{acin2007}, multi-partite entanglement \cite{choi2010, jungnitsch2011} and quantum phase transitions \cite{torma1998, hartmann2006, greentree2006} make quantum networks a prime theme of current theoretical and experimental research. Similarly, quantum networks form the basis of quantum communication and distributed quantum information processing architectures, with interactions taking the form of quantum logic gates \cite{kimble2008, cirac1997, duan2001, briegel1998}.

The physical implementation of quantum networks requires suitable channels and nodes. Photonic channels are well-advanced transmitters of quantum information. Optical photons can carry quantum information over long distances with almost negligible decoherence and are compatible with existing telecommunication fibre technology. The versatility of quantum networks, however, is largely defined by the capability of the network nodes. Dedicated tasks like quantum key distribution can already be achieved using send-only emitter nodes and receive-only detector nodes \cite{eisaman2011}. However, in order to fully exploit the capabilities of quantum networks, functional network nodes are required which are able to send, receive and store quantum information reversibly and efficiently.

The implementation and connection of quantum nodes is a major challenge and different approaches are currently being pursued. An intensely studied example are ensembles of gas-phase atoms \cite{lvovsky2009,hammerer2010,sangouard2011}, but the protocols for the generation of single excitations are inherently probabilistic \cite{duan2001}. Another strong contender are single particles \cite{duan2010}, which allow for single-photon emission \cite{lounis2005}, quantum gate operations \cite{jaksch2000, isenhower2010, timoney2011}, and scalability \cite{home2009}. But single emitters generally exhibit weak light-matter interaction resulting, again, in inherently probabilistic information exchange and very low success rates. In particular, the reversible quantum state mapping between a photonic channel and a single emitter in free space is highly inefficient. In their seminal work \cite{cirac1997}, Cirac and coworkers therefore proposed to overcome these problems by network nodes based on single emitters embedded in optical cavities.

Here we present the first experimental realization of this prototype of a quantum network. The nodes are formed by single atoms quasi-permanently trapped in optical cavities. The cavity-enhanced light-matter interaction opens up a deterministic path for interconversion of photonic and atomic quantum states. By dynamic control of coherent dark states \cite{cirac1997, wilk2007, boozer2007, muecke2010, specht2011}, single photons are reversibly exchanged between distant network nodes. We demonstrate faithful quantum state transfer across a network channel and create entanglement between distant nodes. High fidelities and long coherence times are achieved by encoding the quantum information in the polarization of the photon and the atomic ground-state spin. Our results present the first direct photonic link between two distant single emitters and pave the way for the realization of large-scale quantum networks.

\begin{figure*}
\centering
\includegraphics[width=\textwidth]{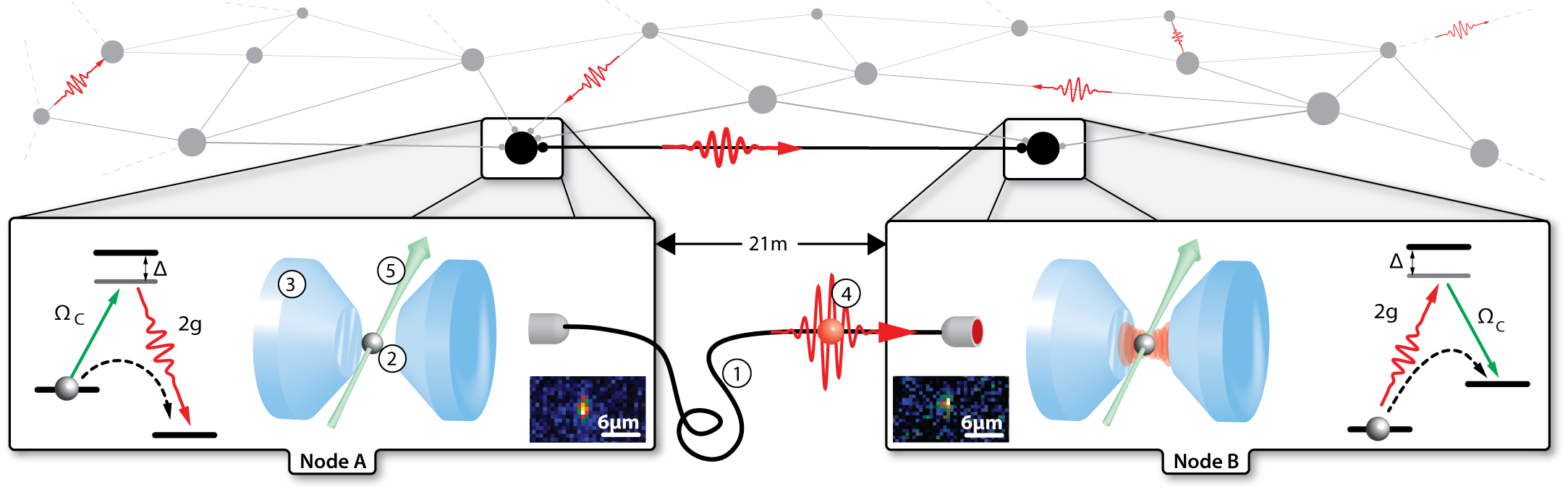}
\caption{\label{fig:setup}
\textbf{A cavity-based quantum network.} In the envisioned architecture, many single-atom nodes are connected by single-photon links. Here we explore the universal properties of connecting two nodes within this configuration. In our experiment these two identical nodes are located in independent laboratories connected by 60\,m optical fibre (1). Each node consists of a single rubidium atom (2) trapped in an optical dipole trap at the centre of a high-finesse optical cavity (3). Quantum state transfer between the atoms and remote entanglement can be achieved via exchange of a single photon (4), with the quantum information encoded in the internal state of the atom and the polarization of the photon. Both, the production of a photon (node A) and its storage (node B) are achieved via a coherent and reversible stimulated Raman adiabatic passage (see main text for details; red: exchanged photon; green and (5): control laser). The inner and outer insets show fluorescence images of the two single atoms and the atomic level scheme, respectively.}
\end{figure*}

In the following, we describe experiments in which we characterize a single network node and the connection of two nodes forming an elementary network. The two nodes are operated in independent laboratories at a distance of 21\,m and are connected by an optical fibre link of 60\,m length. In a first experiment, we demonstrate that single photons can be stored in and retrieved from a single-atom node preserving the photonic polarization state. Second, we show the faithful transfer of arbitrary atomic quantum states from one node to the other. Third, a maximally entangled state of the distant atoms is created with a fidelity of up to 98\,\% and is maintained for at least 100\,\textmu s. This coherence time exceeds the entanglement distribution time across the network link by two orders of magnitude and translates into a maximum possible entangled node distance of 20\,km optical fibre path. Finally, local unitary operations are performed on one of the nodes resulting in rotations of the nonlocal bipartite state whereby different maximally entangled states are created.

\subsection*{A universal quantum network node}
\begin{figure}
\centering
\includegraphics[width=\columnwidth]{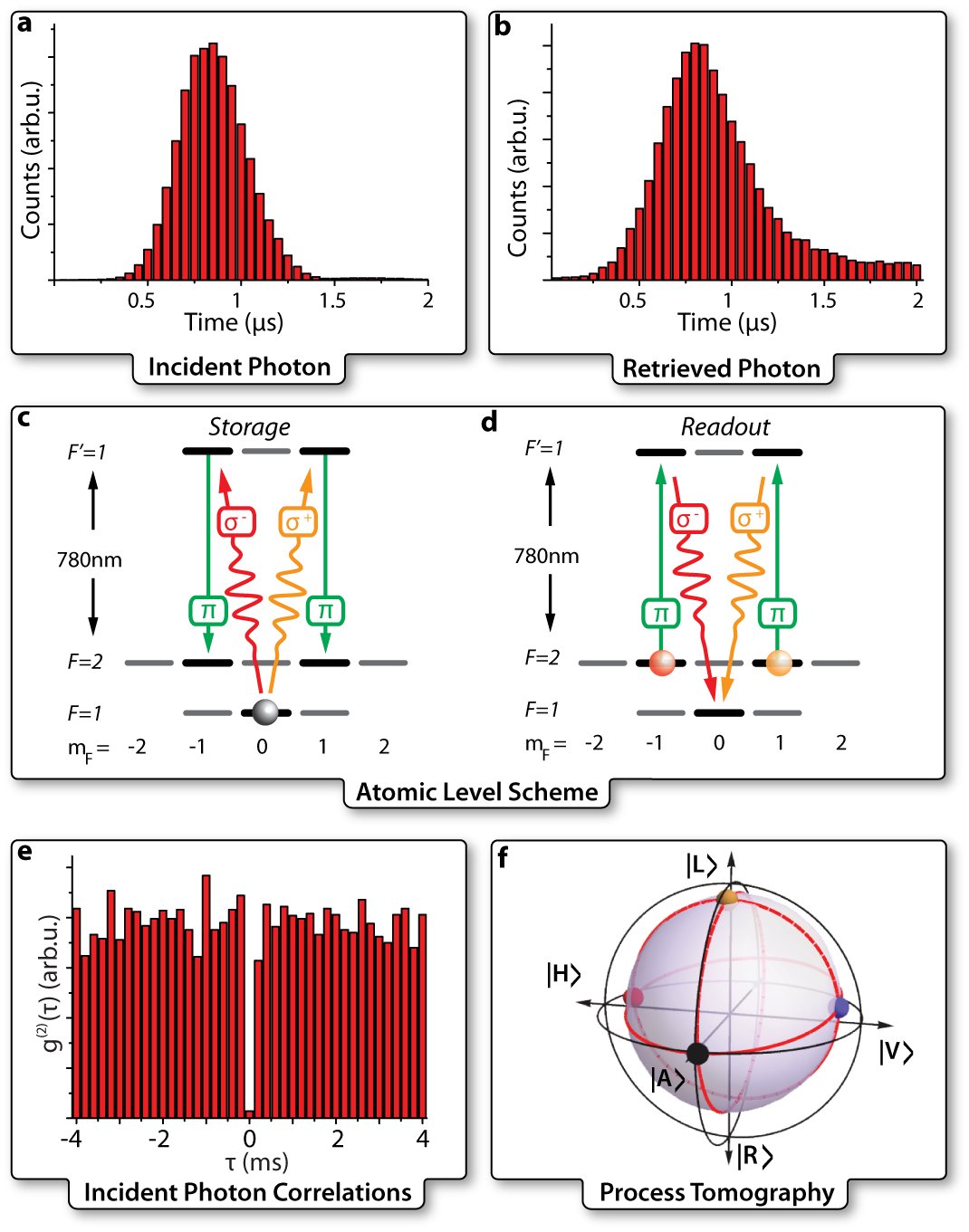}
\caption{\label{fig:single_photon_QM}
\textbf{Universal quantum network node.} \textbf{a}, Single photons with the shown temporal shape (full width at half maximum 0.5\,\textmu s) are produced from node A and subsequently stored in node B.  \textbf{b} They are retrieved after a storage time of 2.5\,\textmu s. \textbf{c}, Optical transitions and atomic states of $^{87}$Rb used in the protocol. The atom at node B is initially prepared in the state $|F=1,m_F=0\rangle$. The polarization of the incoming photon is a superposition of $\sigma^+$ and $\sigma^-$ components (red and yellow arrows) and is converted into a superposition of the $|F=2,m_F=\pm1\rangle$ states via a stimulated Raman adiabatic passage using a $\pi$-polarized control laser (green arrows). In the process, the phase relation between the $\sigma^{\pm}$ polarization components is mapped to a relative phase of the atomic Zeeman states. \textbf{d}, The photon is recreated by reversing the storage process. \textbf{e}, The measured antibunching confirms that single photons are produced by node A. \textbf{f}, Quantum tomography of the storage process. The minimal deformation of the unit Poincaré sphere proves that every initial photonic quantum state is well preserved during storage and retrieval. The fidelity averaged over all input states is $F_\mathrm{qm}=(92.2\pm0.4)$\,\%. All quoted errors are statistical.}
\end{figure}

Each node in our quantum network consists of a single neutral rubidium atom which is quasi-permanently trapped in a high-finesse optical resonator \cite{lettner2011,specht2011} (see Methods). The physical parameters put our atom-cavity systems in the intermediate coupling regime of cavity QED, with the resonators optimized for highly directional optical output into a single mode ($\geq90$\,\%) which is efficiently coupled (up to 90\,\%) to the fibre link (Fig.\,\ref{fig:setup}). The reversible conversion between quantum states of light and matter is enabled by the dynamic control of coherent dark states. The $5^2S_{1/2}$ hyperfine ground states $F=1$ and $F=2$ of the single $^{87}$Rb atom are coupled in a Raman configuration formed by the vacuum mode of the cavity (vacuum Rabi frequency $2g$) and an external control laser field (Rabi frequency $\Omega_C$) (Fig.\,\ref{fig:setup}). Control laser and cavity are both blue-detuned by several tens of MHz from the transition to the excited $5^2P_{3/2}$, $F'=1$ state which -- for an ideal coherent dark state -- is never populated. The atom is driven coherently between the two ground states by controlling $\Omega_C(t)$. With the atom initially prepared in $|F=2,m_F=0\rangle$ and $\Omega_C(0)=0$, a single photon is generated in the cavity mode by increasing the control laser power to $\Omega_C>2g$. Due to the finite resonator decay time the single photon is immediately emitted into the output mode with its temporal wave-packet shape determined by $\Omega_C (t)$ (Fig.\,\ref{fig:single_photon_QM})\cite{kuhn2002}. The single-photon character is confirmed by a second-order correlation measurement (Fig.\,\ref{fig:single_photon_QM}e). The efficiency of the emission process is up to 60\,\%. The inverse process of coherent absorption of a single photon is performed as a time-reversal of the emission process and requires decreasing $\Omega_C (t)$ timed with the arrival of a single photon.

We produce single photons with nearly time-symmetric shape in one atom-cavity system (Fig.\,\ref{fig:single_photon_QM}a) and coherently absorb them in the second system \cite{cirac1997}. After a selectable storage time, the photon is reemitted (Fig.\,\ref{fig:single_photon_QM}b). The overall write-read success rate is typically $(10\pm1)$\,\% (all quoted errors are statistical). With a photon production efficiency of 60\,\%, the storage efficiency is calculated to be 17\,\%. The maximum efficiencies for photon absorption and emission are set by the atom-cavity coupling strength $g$ and can approach unity for smaller cavity mode volumes and vanishing scattering losses at the cavity mirrors.

For encoding one bit of quantum information (qubit), we utilize the photonic polarization degree of freedom and the atomic ground-state spin, i.e. the Zeeman state manifold in each hyperfine state (Fig.\,\ref{fig:single_photon_QM}c and d). When applying a $\pi$-polarized control laser field, the selection rules for electromagnetic dipole transitions ensure the faithful mapping of the polarization state of an incoming photon onto a well-defined superposition of atomic Zeeman states and vice versa for the reemitted photon. To characterize the conversion process, we set the polarization of the incoming photon and compare it to that of the retrieved photon after storage in the atom \cite{specht2011}. The reconstructed Poincaré sphere after a storage time of 2.5\,\textmu s is shown in Fig.\,\ref{fig:single_photon_QM}f. The average fidelity of the quantum memory, defined as the average overlap with the input state, is $F_\mathrm{qm}=(92.2\pm0.4)$\,\% for photons arriving in a 1\,\textmu s time interval (starting at $t=0.2$\,\textmu s in the graph in Fig.\,\ref{fig:single_photon_QM}a, see Discussion). The measured fidelity is far above the classical limit \cite{massar1995} of $2/3$. This experiment is the first to prove coherent transfer of a qubit encoded in a single photon onto a single atom.

\subsection*{Quantum state transfer between single atoms}
\begin{figure*}
\centering
\includegraphics[width=\textwidth]{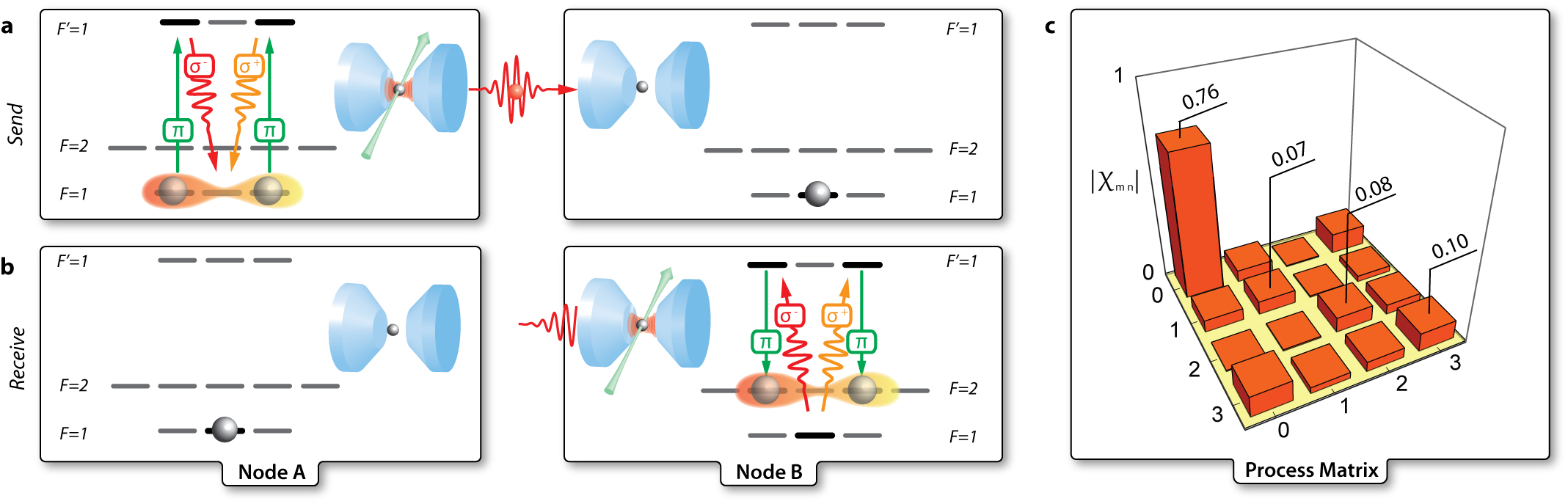}
\caption{\label{fig:state_transfer}
\textbf{Quantum state transfer between two single-atom network nodes.} \textbf{a}, At node A, an arbitrary quantum state is encoded in the Zeeman state manifold of the single atom (see Methods). This quantum information is mapped onto the polarization of a single photon which is sent to node B. \textbf{b}, The photonic polarization is mapped to a superposition of atomic Zeeman states, thereby completing the quantum state transfer from node A to node B. \textbf{c}, Absolute value of the elements of the process matrix $\chi$ for the quantum state transfer. The average fidelity between the ideal and the read-out transferred state is $(84\pm1)$\,\%, well above the classical limit of $2/3$.}
\end{figure*}

In the next experiment, we transfer quantum information from node A to the distant node B by sending a single photon across the fibre link. The qubit to be transferred is the state of the atom in node A represented by
$$|\psi_\mathrm{A}\rangle = \alpha |F=1,m_F=-1\rangle+\beta |F=1,m_F=+1\rangle,$$
where $\alpha$ and $\beta$ are normalized complex-valued amplitudes. We apply a $\pi$-polarized control laser pulse to the atom at node A, thereby generating a photon in the polarization state
$$|\psi_\mathrm{photon}\rangle = \alpha |L\rangle+\beta |R\rangle.$$
Here $|L\rangle$ and $|R\rangle$ refer to the left and right circular polarization components of the photon. After emission of the photon, the atom at node A is left in the state $|F=1,m_F=0\rangle$ (Fig.\,\ref{fig:state_transfer}a, left diagram and ref.\,\cite{wilk2007}). The photon is transmitted through the optical fibre to node B. The atom at node B is initially prepared in the state $|F=1,m_F=0\rangle$. Using the Raman scheme (Fig.\,\ref{fig:state_transfer}b, right diagram) the incoming photon is absorbed and its polarization state is mapped onto the atomic state of node B, which becomes
$$|\psi_\mathrm{B}\rangle = \alpha |F=2,m_F=-1\rangle+\beta |F=2,m_F=+1\rangle.$$
After absorption, an arbitrary quantum state has been successfully communicated from node A to node B. The qubit encoded in atom B is identical to the original one in node A, albeit encoded in the $F=2$ hyperfine manifold. After this quantum state transfer, node A is now ready to receive another photonic qubit, whereas node B is capable of resending the stored qubit at any time. It is this symmetric and reversible feature that makes this scheme scalable to arbitrary network configurations of multiple atom-cavity systems.

We analyse the quantum state transfer using quantum process tomography \cite{nielsen2000}. For this purpose, we prepare the atom at node A in a state of the form $|\psi_\mathrm{A}\rangle$ by a projective measurement (see Methods). After quantum state transfer from node A to node B as described above, we read out the state of the atom in node B by mapping it onto the polarization of a second single photon which is then detected. By comparing a sufficient set of initial quantum states in node A with the obtained states in node B we can infer the outcome of the protocol for any initial state of node A. The process matrix $\chi$ describes the mapping of the density matrix $\rho_\mathrm{A}$ of the state at node A (taken to be the ideally prepared state $|\psi_\mathrm{A}\rangle$) onto the transferred state $\rho_\mathrm{B}$ at node B through the operation $\rho_\mathrm{B} = \sum_{m,n=0}^3 \chi_{mn}\sigma_m \rho_\mathrm{A} \sigma_n^\dagger$. Here the $\sigma_i$ are a set of operators given by the Pauli matrices. For calculating $\chi$ we normalize the density matrices.

Ideally, the two density matrices are identical, which is equivalent to having $\chi_{00}=1$ and all other elements zero. Fig.\,\ref{fig:state_transfer}c shows the absolute value of all elements of $\chi$ obtained from a maximum likelihood fit to the experimental data. We find $\chi_{00}=0.76$ as the dominating matrix element indicating a high level of control over the quantum process. The main deviation from a perfect state transfer is a slight depolarization of the quantum state, as indicated by the non-vanishing diagonal elements of $\chi_{11}$, $\chi_{22}$ and $\chi_{33}$. The state transfer can also be characterized by a fidelity defined as the average overlap between initial and transferred state. We find $F_\mathrm{st}=(84\pm1.0)$\,\%, which proves the quantum character of the state transfer, as the highest fidelity achievable with classical information exchange between the nodes is $2/3$. The overall success rate of the state transfer protocol is 0.2\,\%, resulting from a production efficiency of the transmitter photon at node A of 3\,\% (see Methods), propagation losses leading to a photon transmission of 34\,\% and a storage efficiency at node B of about 20\,\%.

\subsection*{Remote atom-atom entanglement}
The most remarkable property of a quantum network is the existence of entangled quantum states shared among several network nodes. This is a basis for quantum logic gate operations between nodes and can lead to complex quantum many-body phenomena.

\begin{figure}
\centering
\includegraphics[width=\columnwidth]{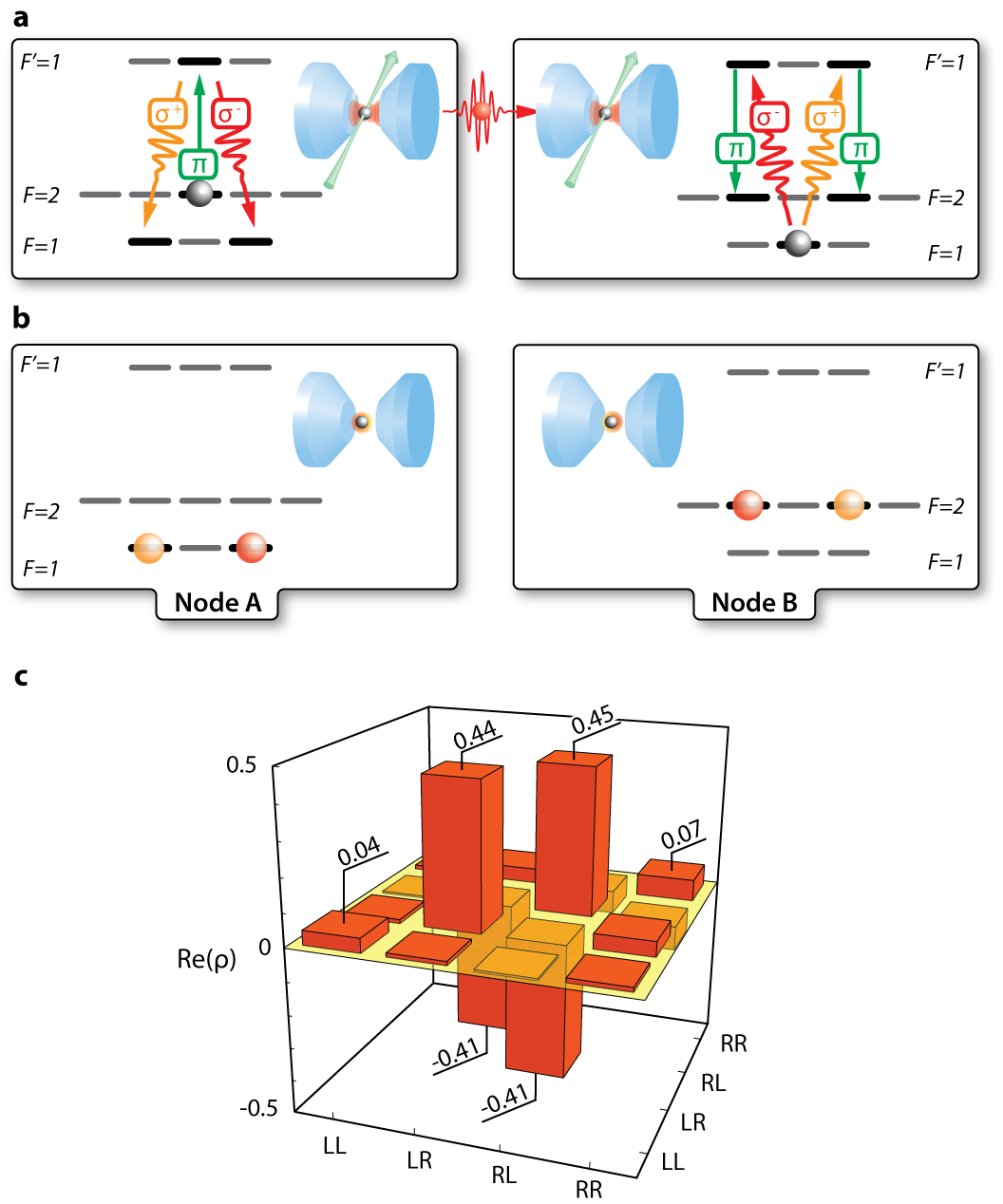}
\caption{\label{fig:remote_entanglement}
\textbf{Remote entanglement of two single-atom nodes.} \textbf{a}, A single photon is generated at node A, such that the internal state of the atom and the polarization of the photon are entangled. The photon is sent to node B where its polarization is mapped onto the atomic state. \textbf{b} This creates entanglement between nodes A and B that can be maintained for at least 100\,\textmu s. \textbf{c}, For analysis, both atomic states are converted into single photons. Polarization tomography on the two photons confirms the entanglement between the two nodes. We measure a fidelity of $F_{|\Psi^-\rangle}=(85\pm1.3)$\,\% with respect to the $|\Psi^-\rangle$ Bell state. Shown is the real part of the density matrix. The magnitude of each imaginary parts is $\leq0.03$.}
\end{figure}

In the following, we demonstrate the creation of remote entanglement between distant single-atom nodes based on the transmission of a single photon.
We first prepare the atom at node A in the state $|F=2,m_F=0\rangle$ (Fig.\,\ref{fig:remote_entanglement}). Applying a $\pi$-polarized control laser pulse triggers the emission of a single photon and creates the maximally entangled state
$$|\psi_{\rm A\otimes photon}^-\rangle = \frac{1}{\sqrt{2}}\left(|1,-1\rangle\otimes|R\rangle-|1,1\rangle\otimes|L\rangle\right)$$
between the spin state of the atom and the polarization of the photon \cite{wilk2007} that is routed to node B. There it is coherently absorbed and its polarization is mapped onto the spin state of atom B. The atom-photon entanglement is thus converted into entanglement between the two nodes, with the two atoms in the maximally entangled $|\Psi^-\rangle$ Bell state
$$|\psi_{\rm A\otimes B}^-\rangle = \frac{1}{\sqrt{2}}\left(|1,-1\rangle\otimes|2,1\rangle-|1,1\rangle\otimes|2,-1\rangle\right).$$

We verify the presence of this entangled state by mapping the atomic state at each node onto a photon and analyzing the polarization correlations among the two read-out photons. The real part of the resulting density matrix, with the read-out performed 7\,\textmu s after the creation of atom-atom entanglement, is shown in Fig.\,\ref{fig:remote_entanglement}c. We find a fidelity of $F_{|\Psi^-\rangle}=(85\pm1.3)$\,\% with the $|\Psi^-\rangle$ Bell state. This exceeds the classical limit of 50\,\%, clearly proving the existence of entanglement between the two remote atoms. Fidelities as high as $(98.7\pm2.2)$\,\% can be achieved by further post-selection of photon detection events (see Methods).

The success probability of entanglement creation is 2\,\%. It is the product of the photon generation efficiency (40\,\%) at node A, the probability with which the photon is delivered to node B (34\,\%) and its storage efficiency at node B (14\,\%). The verification process for the entanglement, consisting of the production of one photon at each of the two nodes and their subsequent detection, has an efficiency of 0.16\,\%.

The entanglement created in this experiment exists between the spin states of two single atoms at a physical distance of 21\,m. Highly non-classical correlations between the two atoms are observed for 100\,\textmu s. The fidelity with the $|\Psi^-\rangle$ Bell state measured 100\,\textmu s after creation of entanglement is $(56\pm3)$\,\%, still exceeding the classical threshold of 50\,\% by two standard deviations. The decoherence of the atom-atom entangled state is dominated by dephasing caused by uncorrelated magnetic field fluctuations (on the order of 1\,mG) at the two individual nodes and position-dependent differential AC Stark shifts induced by the dipole trap light fields. The dephasing due to local magnetic field fluctuations can be significantly reduced by applying small magnetic guiding fields (30\,mG) along the quantization axis of each node. This has been used for the measurements with 100\,\textmu s entanglement duration. The observed remote-entanglement lifetime exceeds the entanglement creation time (1\,\textmu s for creation, transmission and absorption of an entangling photon) by two orders of magnitude. Entanglement lifetimes on the order of seconds can be expected when mapping the Zeeman qubit to magnetic-field-insensitive clock states using microwave or Raman pulses \cite{olmschenk2007}.

In the limit of unit efficiency, the entanglement scheme presented here allows for the deterministic creation of entanglement. In our experimental implementation, efficiencies are below one and we therefore detect \textit{a posteriori} entanglement \cite{enk2007}. The detection of entangled read-out photons indicates that atom-atom entanglement had been present. Only entanglement attempts that lead to the final detection of two read-out photons in the mapping process are considered in our data. The creation of \textit{heralded} entanglement is possible by implementing a mechanism that signals the successful storage of a transmitted photon at node B without disturbing the stored quantum state. This can be realized by hyperfine state detection using $F=1$ as the bright state \cite{lloyd2001,bochmann2010,volz2011}.

\subsection*{Local manipulation of a nonlocal state}
\begin{figure}
\centering
\includegraphics[width=\columnwidth]{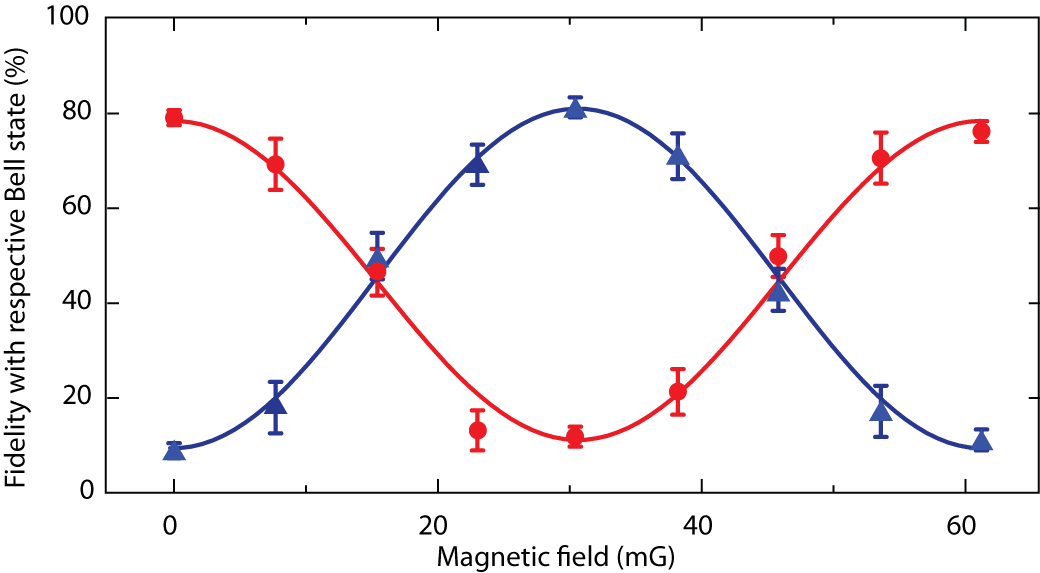}
\caption{\label{fig:state_rotation}
\textbf{Controlled rotation of the entangled state.} A magnetic field, locally applied at node B, changes the non-local quantum state of the entangled nodes. The atom at node A is held at zero magnetic field. The phase evolution for a fixed hold time of 12.5\,\textmu s is proportional to the applied magnetic field. Plotted are the fidelity with the $|\Psi^-\rangle$ and $|\Psi^+\rangle$ Bell state (red and blue data, respectively), showing a characteristic oscillation. The solid line is a cosine fit to guide the eye. The initially prepared $|\Psi^-\rangle$ state ($F_{|\Psi^-\rangle}=(79\pm1.6)$\,\%) can be rotated into a $|\Psi^+\rangle$ state of comparable fidelity ($F_{|\Psi^+\rangle}=(81\pm2)$\,\%) using a magnetic field of 30\,mG.}
\end{figure}

Nodes A and B are in separated physical locations and thus are independently addressable for local qubit control. When two nodes are entangled, unitary operations applied locally at one of the nodes change the nonlocal state of both nodes while the entanglement is preserved. Thus local qubit control allows one to create arbitrary maximally entangled two-qubit states using a single initial entangled state as a resource. We demonstrate this capability by creating the $|\Psi^+\rangle$ Bell state. We start by preparing the two nodes in the $|\Psi^-\rangle$ Bell state as described above. Applying a magnetic field along the quantization axis only at node B causes a state rotation at twice the Larmor frequency. The fidelity of the created state with the $|\Psi^-\rangle$ and the $|\Psi^+\rangle$ Bell state is plotted as a function of the applied magnetic field in Fig.\,\ref{fig:state_rotation}. The time between entanglement creation and read-out of the atomic state is fixed at 12.5\,\textmu s. As can be seen from Fig.\,\ref{fig:state_rotation}, the rotation of the nonlocal state results in a sinusoidally varying overlap with the $|\Psi^\pm\rangle$ Bell states. The fidelity with respect to the $|\Psi^+\rangle$ state reaches a maximum of $(81\pm2)$\,\% at a magnetic field of $B=30$\,mG. The original $|\Psi^-\rangle$ state is recovered with a fidelity of $(76\pm2)$\,\% after a spin rotation of $2\pi$. The reduced fidelity with the $|\Psi^-\rangle$ state at a field of 60\,mG is a result of non-negligible Larmor precession during the entanglement creation and read-out processes.

\subsection*{Discussion}
The body of work presented here constitutes the first direct coupling of two distant single quantum emitters by exchange of a single photon. Our results introduce universal quantum network nodes based on single emitters. Single-atom--cavity nodes excel previously investigated light-matter interfaces and incorporate their specific advantages in one platform. The use of single emitters offers a clear perspective for heralding \cite{lloyd2001,bochmann2010,volz2011} and the integration of quantum gate operations both local and remote \cite{jaksch2000, isenhower2010, timoney2011, duan2010, duan2005}. In the following, we briefly discuss the potential of our specific implementation of universal quantum network nodes with respect to fidelity and efficiency of the described processes, storage time, and scalability.

In all our experiments, atomic state preparation errors due to non-optimal optical pumping reduce the fidelity. Not perfectly $\pi$-polarized control lasers and off-resonant excitations cause deviations from the ideal transition scheme (Fig.\,\ref{fig:single_photon_QM}, \ref{fig:state_transfer} and \ref{fig:remote_entanglement}). These errors may lead to emission of photons with excitation paths different from the Raman scheme. A detailed analysis of transition strengths and effective Rabi frequencies shows that these excitation paths lead to delayed photon emission. The contribution of these photons to the measured signal can therefore be suppressed by post-selecting subsets of data based on photon arrival times (see Methods). As an example, entangled-state fidelities as high as $(98.7\pm2.2)$\,\% are reached when considering only the first 50\,\% of read-out photons from node A and the initial 14\,\% of photons retrieved from node B. These results show the great potential for achieving very high fidelities if the mentioned imperfections are overcome.

The efficiency achievable with the demonstrated deterministic entanglement scheme \cite{cirac1997} is higher than what can be achieved with probabilistic schemes \cite{duan2001,duan2010}. In our experiments, it is largely limited by the moderate atom-cavity coupling strength $g$. Efficiencies approaching unity may be accomplished when the cavity mode volume is decreased \cite{volz2011} and mirror scattering losses are eliminated. The very weak coupling of the nuclear spin of single atoms to the environment can be exploited to boost the coherence times of our network nodes by mapping the Zeeman qubit onto magnetic-field-insensitive clock states \cite{olmschenk2007}.

The number of qubits per node may be increased through the use of optical lattices and single atom registers \cite{nussmann2005}. After the preparation of remote atom-atom entanglement akin to the procedure shown here, the registers in different nodes could be shifted to successively produce many sets of entangled atoms which can then be used, e.g. for nested entanglement purification. This possibility, in combination with the long storage times achievable with single atoms, represents a realistic avenue towards quantum communication over arbitrary distances by means of a quantum repeater protocol \cite{briegel1998}.

Current advances in photonic technologies allow reconfigurable routing between different nodes, thereby enabling various different network topologies. The controlled interaction between arbitrary nodes and the plethora of accessible topologies of cavity quantum networks are not only an important resource for quantum information processing---cavity networks also constitute a suitable paradigm for investigating emergent phenomena such as quantum phase transitions of light \cite{torma1998,hartmann2006,greentree2006} or percolation of entanglement \cite{acin2007}.

\begin{acknowledgments}
We thank David Moehring for his contributions during the early stage of the experiments and Benedikt Mayer and Milan Padilla from the Walter Schottky Institut for gold coating of the fast-moving mirror. This work was supported by the Deutsche Forschungsgemeinschaft (Research Unit 635), by the European Union (Collaborative Project AQUTE) and by the Bundesministerium für Bildung und Forschung via IKT 2020 (QK\_QuOReP). E.\,F. acknowledges support from the Alexander von Humboldt Foundation.\\
\end{acknowledgments}

\section*{Methods}

\subsection*{Experimental set-up}
The two independent quantum nodes are designed to operate with similar physical parameters. In each apparatus, a single $^{87}$Rb atom is quasi-permanently trapped inside an optical dipole trap (potential depth $U_0/k_B=3$ and 5\,mK) and held at the centre of a high-finesse optical cavity (finesse $6\times10^4$, mirror distance 0.5\,mm, mode waist 30\,\textmu m). Both cavities have asymmetric mirror transmissions of $T_1<6$\,ppm and $T_2\approx 100$\,ppm, leading to a highly directional ($\ge0.9$) single output mode which is matched to a single-mode optical fibre (efficiency up to 0.9). A fast-moving mirror can switch the network from a configuration with optically connected nodes to a configuration in which the light emitted from the two nodes is guided to two polarization detection setups. A single photon generated inside one of the cavities is eventually detected with an efficiency of 0.3, owing to the quantum efficiency of the detector (0.6) and all propagation losses. Both systems produce photons on the D$_2$ line of $^{87}$Rb at a wavelength of 780\,nm. In this configuration both atom-cavity systems operate in the intermediate coupling regime of cavity QED (coherent atom-cavity coupling $g\leq 2\pi\times5$\,MHz, cavity field decay rate $\kappa=2\pi\times3$\,MHz, atomic polarization decay rate $\gamma=2\pi\times3$\,MHz). When a single atom is trapped inside each of the cavities, the experimental protocol runs at a repetition rate of 5\,kHz, including optical pumping (20\,\textmu s), photon generation (1\,\textmu s), photon storage (1--100\,\textmu s) and optical cooling of atomic motion (80\,\textmu s). The necessary laser beams impinge perpendicular to the cavity axis. The presence and position of single atoms is monitored in realtime by an electron multiplying CCD camera, which collects atomic fluorescence light (inset in Fig.\,\ref{fig:setup}). In combination with a longitudinally shiftable standing-wave dipole trap, the atoms are actively positioned at the centre of the cavity mode \cite{nussmann2005}. Single-atom storage times are on the order of one minute.

\subsection*{Projective atomic state preparation for quantum state transfer}
To characterize the quality of the quantum state transfer from node A to node B, the atom at node A is prepared in one of six different initial states of the form $|\psi_\mathrm{A}\rangle = \alpha|F=1,m_F=-1\rangle+\beta|F=1,m_F=+1\rangle,$ forming a regular octahedron on the Poincaré sphere. This is achieved through a projective measurement. The atom is initialized in the state $|F=2,m_F=0\rangle$, and subsequently a single photon is generated using a $\pi$-polarized laser pulse. This creates the atom-photon entangled state $|\psi_{\rm A\otimes photon}\rangle = \frac{1}{\sqrt{2}}\left(|1,-1\rangle\otimes|R\rangle-|1,1\rangle\otimes|L\rangle\right)$. Detection of this photon in a well-defined polarization state projects the atom in node A onto a qubit state $|\psi_\mathrm{A}\rangle$, with ($\alpha, \beta$) determined by the particular choice of the detector's polarization basis. Following this projective preparation, the known quantum state of node A is transferred to node B. Read-out of the state of the atom at node B is performed by mapping it onto a single photon whose polarization can be analysed. In calculating the fidelity of the quantum process, we assume perfect preparation of $|\psi_\mathrm{A}\rangle$. It is therefore a lower bound on the fidelity of the state transfer.

\subsection*{Post-selected fidelities}
The ideal Raman schemes depicted in Figs.\,\ref{fig:single_photon_QM}, \ref{fig:state_transfer} and \ref{fig:remote_entanglement} lead to emission of spatio-temporally well-defined single-photon wave packets with their polarization determined by the selection rules. The fidelities reported in the previous sections are obtained from analyzing correlations between photon detection events in different polarization bases. We have identified several experimental imperfections which cause deviations from this ideal Raman scheme: imperfect initial state preparation, misaligned polarization of the control laser and off-resonant excitations. These imperfections not only affect polarization correlations but also the temporal wave packet shape of the emitted photons. A detailed analysis has shown that these imperfections are generally correlated with delayed photon emission with respect to the ideal Raman scheme. In this Article, we usually evaluate all read-out photons from node A and those photons from node B arriving within a 1\,\textmu s time interval centred around the maximum of the photon wave packet (see Fig.\,\ref{fig:single_photon_QM}a). The contribution of the mentioned non-ideal processes to the measured fidelities can however be minimized when only those photon detection events are analysed that occur early in the photon's temporal wave packet. Indeed, we find close to ideal fidelities for these subsets of data. In the atom-atom entanglement experiment the fidelity is increased to $F_{|\Psi^-\rangle}=(98.7\pm2.2)$\,\% when considering only the first 50\,\% of the ensemble of detected photons from node A and the initial 14\,\% of the ensemble of photons from node B. The quoted values for $F_{|\Psi^-\rangle}$ are the unbiased estimator and the statistical standard error. The likelihood function of $F_{|\Psi^-\rangle}$ is non-Gaussian.

The imperfections correlated with late photon emission can also be suppressed by tailoring the single-photon read-out process directly. When the applied control laser Rabi frequencies $\Omega_C$ are kept low and applied for only a short time, the read-out photon can be made to resemble the post-selected subset. We have made use of this weak read-out in the experiments on state transfer and remote atom-atom entanglement, thereby optimizing for high fidelities at the expense of the efficiency of the process. As all presented schemes are intrinsically deterministic, future improvements on the imperfect processes mentioned above will allow for both, fidelities and efficiencies approaching unity, without the necessity for any tradeoff.


\begin{thebibliography}{10}
\expandafter\ifx\csname url\endcsname\relax
  \def\url#1{\texttt{#1}}\fi
\expandafter\ifx\csname urlprefix\endcsname\relax\def\urlprefix{URL }\fi
\providecommand{\bibinfo}[2]{#2}
\providecommand{\eprint}[2][]{\url{#2}}

\bibitem{acin2007}
\bibinfo{author}{Acín, A.}, \bibinfo{author}{Cirac, J.~I.} \&
  \bibinfo{author}{Lewenstein, M.}
\newblock \bibinfo{title}{Entanglement percolation in quantum networks}.
\newblock \emph{\bibinfo{journal}{Nature Phys.}} \textbf{\bibinfo{volume}{3}},
  \bibinfo{pages}{256--259} (\bibinfo{year}{2007}).

\bibitem{choi2010}
\bibinfo{author}{Choi, K.~S.}, \bibinfo{author}{Goban, A.},
  \bibinfo{author}{Papp, S.~B.}, \bibinfo{author}{van Enk, S.~J.} \&
  \bibinfo{author}{Kimble, H.~J.}
\newblock \bibinfo{title}{Entanglement of spin waves among four quantum
  memories}.
\newblock \emph{\bibinfo{journal}{Nature}} \textbf{\bibinfo{volume}{468}},
  \bibinfo{pages}{412--416} (\bibinfo{year}{2010}).

\bibitem{jungnitsch2011}
\bibinfo{author}{Jungnitsch, B.}, \bibinfo{author}{Moroder, T.} \&
  \bibinfo{author}{Gühne, O.}
\newblock \bibinfo{title}{Taming multiparticle entanglement}.
\newblock \emph{\bibinfo{journal}{Phys. Rev. Lett.}}
  \textbf{\bibinfo{volume}{106}}, \bibinfo{pages}{190502}
  (\bibinfo{year}{2011}).

\bibitem{torma1998}
\bibinfo{author}{Törmä, P.}
\newblock \bibinfo{title}{Transitions in quantum networks}.
\newblock \emph{\bibinfo{journal}{Phys. Rev. Lett.}}
  \textbf{\bibinfo{volume}{81}}, \bibinfo{pages}{2185--2189}
  (\bibinfo{year}{1998}).

\bibitem{hartmann2006}
\bibinfo{author}{Hartmann, M.~J.}, \bibinfo{author}{Brand\~{a}o, F. G. S.~L.}
  \& \bibinfo{author}{Plenio, M.~B.}
\newblock \bibinfo{title}{Strongly interacting polaritons in coupled arrays of
  cavities}.
\newblock \emph{\bibinfo{journal}{Nature Phys.}} \textbf{\bibinfo{volume}{2}},
  \bibinfo{pages}{849--855} (\bibinfo{year}{2006}).

\bibitem{greentree2006}
\bibinfo{author}{Greentree, A.~D.}, \bibinfo{author}{Tahan, C.},
  \bibinfo{author}{Cole, J.~H.} \& \bibinfo{author}{Hollenberg, L. C.~L.}
\newblock \bibinfo{title}{Quantum phase transitions of light}.
\newblock \emph{\bibinfo{journal}{Nature Phys.}} \textbf{\bibinfo{volume}{2}},
  \bibinfo{pages}{856--861} (\bibinfo{year}{2006}).

\bibitem{kimble2008}
\bibinfo{author}{Kimble, H.~J.}
\newblock \bibinfo{title}{The quantum internet}.
\newblock \emph{\bibinfo{journal}{Nature}} \textbf{\bibinfo{volume}{453}},
  \bibinfo{pages}{1023--1030} (\bibinfo{year}{2008}).

\bibitem{cirac1997}
\bibinfo{author}{Cirac, J.~I.}, \bibinfo{author}{Zoller, P.},
  \bibinfo{author}{Kimble, H.~J.} \& \bibinfo{author}{Mabuchi, H.}
\newblock \bibinfo{title}{Quantum state transfer and entanglement distribution
  among distant nodes in a quantum network}.
\newblock \emph{\bibinfo{journal}{Phys. Rev. Lett.}}
  \textbf{\bibinfo{volume}{78}}, \bibinfo{pages}{3221--3224}
  (\bibinfo{year}{1997}).

\bibitem{duan2001}
\bibinfo{author}{Duan, L.-M.}, \bibinfo{author}{Lukin, M.~D.},
  \bibinfo{author}{Cirac, J.~I.} \& \bibinfo{author}{Zoller, P.}
\newblock \bibinfo{title}{Long-distance quantum communication with atomic
  ensembles and linear optics}.
\newblock \emph{\bibinfo{journal}{Nature}} \textbf{\bibinfo{volume}{414}},
  \bibinfo{pages}{413--418} (\bibinfo{year}{2001}).

\bibitem{briegel1998}
\bibinfo{author}{Briegel, H.-J.}, \bibinfo{author}{Dür, W.},
  \bibinfo{author}{Cirac, J.~I.} \& \bibinfo{author}{Zoller, P.}
\newblock \bibinfo{title}{Quantum repeaters: The role of imperfect local
  operations in quantum communication}.
\newblock \emph{\bibinfo{journal}{Phys. Rev. Lett.}}
  \textbf{\bibinfo{volume}{81}}, \bibinfo{pages}{5932--5935}
  (\bibinfo{year}{1998}).

\bibitem{eisaman2011}
\bibinfo{author}{Eisaman, M.~D.}, \bibinfo{author}{Fan, J.},
  \bibinfo{author}{Migdall, A.} \& \bibinfo{author}{Polyakov, S.~V.}
\newblock \bibinfo{title}{Invited review article: Single-photon sources and
  detectors}.
\newblock \emph{\bibinfo{journal}{Rev. Sci. Instrum.}}
  \textbf{\bibinfo{volume}{82}}, \bibinfo{pages}{071101}
  (\bibinfo{year}{2011}).

\bibitem{lvovsky2009}
\bibinfo{author}{Lvovsky, A.~I.}, \bibinfo{author}{Sanders, B.~C.} \&
  \bibinfo{author}{Tittel, W.}
\newblock \bibinfo{title}{Optical quantum memory}.
\newblock \emph{\bibinfo{journal}{Nature Photon.}}
  \textbf{\bibinfo{volume}{3}}, \bibinfo{pages}{706--714}
  (\bibinfo{year}{2009}).

\bibitem{hammerer2010}
\bibinfo{author}{Hammerer, K.}, \bibinfo{author}{S{\o}rensen, A.~S.} \&
  \bibinfo{author}{Polzik, E.~S.}
\newblock \bibinfo{title}{Quantum interface between light and atomic
  ensembles}.
\newblock \emph{\bibinfo{journal}{Rev. Mod. Phys.}}
  \textbf{\bibinfo{volume}{82}}, \bibinfo{pages}{1041} (\bibinfo{year}{2010}).

\bibitem{sangouard2011}
\bibinfo{author}{Sangouard, N.}, \bibinfo{author}{Simon, C.},
  \bibinfo{author}{de~Riedmatten, H.} \& \bibinfo{author}{Gisin, N.}
\newblock \bibinfo{title}{Quantum repeaters based on atomic ensembles and
  linear optics}.
\newblock \emph{\bibinfo{journal}{Rev. Mod. Phys.}}
  \textbf{\bibinfo{volume}{83}}, \bibinfo{pages}{33--80}
  (\bibinfo{year}{2011}).

\bibitem{duan2010}
\bibinfo{author}{Duan, L.-M.} \& \bibinfo{author}{Monroe, C.}
\newblock \bibinfo{title}{Colloquium: Quantum networks with trapped ions}.
\newblock \emph{\bibinfo{journal}{Rev. Mod. Phys.}}
  \textbf{\bibinfo{volume}{82}}, \bibinfo{pages}{1209--1224}
  (\bibinfo{year}{2010}).

\bibitem{lounis2005}
\bibinfo{author}{Lounis, B.} \& \bibinfo{author}{Orrit, M.}
\newblock \bibinfo{title}{Single-photon sources}.
\newblock \emph{\bibinfo{journal}{Rep. Prog. Phys.}}
  \textbf{\bibinfo{volume}{68}}, \bibinfo{pages}{1129--1179}
  (\bibinfo{year}{2005}).

\bibitem{jaksch2000}
\bibinfo{author}{Jaksch, D.} \emph{et~al.}
\newblock \bibinfo{title}{Fast quantum gates for neutral atoms}.
\newblock \emph{\bibinfo{journal}{Phys. Rev. Lett.}}
  \textbf{\bibinfo{volume}{85}}, \bibinfo{pages}{2208--2211}
  (\bibinfo{year}{2000}).

\bibitem{isenhower2010}
\bibinfo{author}{Isenhower, L.} \emph{et~al.}
\newblock \bibinfo{title}{Demonstration of a neutral atom controlled-{NOT}
  quantum gate}.
\newblock \emph{\bibinfo{journal}{Phys. Rev. Lett.}}
  \textbf{\bibinfo{volume}{104}}, \bibinfo{pages}{010503}
  (\bibinfo{year}{2010}).

\bibitem{timoney2011}
\bibinfo{author}{Timoney, N.} \emph{et~al.}
\newblock \bibinfo{title}{Quantum gates and memory using microwave-dressed
  states}.
\newblock \emph{\bibinfo{journal}{Nature}} \textbf{\bibinfo{volume}{476}},
  \bibinfo{pages}{185--188} (\bibinfo{year}{2011}).

\bibitem{home2009}
\bibinfo{author}{Home, J.~P.} \emph{et~al.}
\newblock \bibinfo{title}{Complete methods set for scalable ion trap quantum
  information processing}.
\newblock \emph{\bibinfo{journal}{Science}} \textbf{\bibinfo{volume}{325}},
  \bibinfo{pages}{1227--1230} (\bibinfo{year}{2009}).

\bibitem{wilk2007}
\bibinfo{author}{Wilk, T.}, \bibinfo{author}{Webster, S.~C.},
  \bibinfo{author}{Kuhn, A.} \& \bibinfo{author}{Rempe, G.}
\newblock \bibinfo{title}{Single-atom single-photon quantum interface}.
\newblock \emph{\bibinfo{journal}{Science}} \textbf{\bibinfo{volume}{317}},
  \bibinfo{pages}{488--490} (\bibinfo{year}{2007}).

\bibitem{boozer2007}
\bibinfo{author}{Boozer, A.~D.}, \bibinfo{author}{Boca, A.},
  \bibinfo{author}{Miller, R.}, \bibinfo{author}{Northup, T.~E.} \&
  \bibinfo{author}{Kimble, H.~J.}
\newblock \bibinfo{title}{Reversible state transfer between light and a single
  trapped atom}.
\newblock \emph{\bibinfo{journal}{Phys. Rev. Lett.}}
  \textbf{\bibinfo{volume}{98}}, \bibinfo{pages}{193601}
  (\bibinfo{year}{2007}).

\bibitem{muecke2010}
\bibinfo{author}{Mücke, M.} \emph{et~al.}
\newblock \bibinfo{title}{Electromagnetically induced transparency with single
  atoms in a cavity}.
\newblock \emph{\bibinfo{journal}{Nature}} \textbf{\bibinfo{volume}{465}},
  \bibinfo{pages}{755--758} (\bibinfo{year}{2010}).

\bibitem{specht2011}
\bibinfo{author}{Specht, H.~P.} \emph{et~al.}
\newblock \bibinfo{title}{A single-atom quantum memory}.
\newblock \emph{\bibinfo{journal}{Nature}} \textbf{\bibinfo{volume}{473}},
  \bibinfo{pages}{190--193} (\bibinfo{year}{2011}).

\bibitem{lettner2011}
\bibinfo{author}{Lettner, M.} \emph{et~al.}
\newblock \bibinfo{title}{Remote entanglement between a single atom and a
  {B}ose-{E}instein condensate}.
\newblock \emph{\bibinfo{journal}{Phys. Rev. Lett.}}
  \textbf{\bibinfo{volume}{106}}, \bibinfo{pages}{210503}
  (\bibinfo{year}{2011}).

\bibitem{kuhn2002}
\bibinfo{author}{Kuhn, A.}, \bibinfo{author}{Hennrich, M.} \&
  \bibinfo{author}{Rempe, G.}
\newblock \bibinfo{title}{Deterministic single-photon source for distributed
  quantum networking}.
\newblock \emph{\bibinfo{journal}{Phys. Rev. Lett.}}
  \textbf{\bibinfo{volume}{89}}, \bibinfo{pages}{067901}
  (\bibinfo{year}{2002}).

\bibitem{massar1995}
\bibinfo{author}{Massar, S.} \& \bibinfo{author}{Popescu, S.}
\newblock \bibinfo{title}{Optimal extraction of information from finite quantum
  ensembles}.
\newblock \emph{\bibinfo{journal}{Phys. Rev. Lett.}}
  \textbf{\bibinfo{volume}{74}}, \bibinfo{pages}{1259--1263}
  (\bibinfo{year}{1995}).

\bibitem{nielsen2000}
\bibinfo{author}{Nielsen, M.~A.} \& \bibinfo{author}{Chuang, I.~L.}
\newblock \emph{\bibinfo{title}{Quantum Computation and Quantum Information}}
  (\bibinfo{publisher}{Cambridge University Press},
  \bibinfo{address}{Cambridge}, \bibinfo{year}{2000}).

\bibitem{olmschenk2007}
\bibinfo{author}{Olmschenk, S.} \emph{et~al.}
\newblock \bibinfo{title}{Manipulation and detection of a trapped {Y}b$^{+}$
  hyperfine qubit}.
\newblock \emph{\bibinfo{journal}{Phys. Rev. A}} \textbf{\bibinfo{volume}{76}},
  \bibinfo{pages}{052314} (\bibinfo{year}{2007}).

\bibitem{enk2007}
\bibinfo{author}{van Enk, S.~J.}, \bibinfo{author}{Lütkenhaus, N.} \&
  \bibinfo{author}{Kimble, H.~J.}
\newblock \bibinfo{title}{Experimental procedures for entanglement
  verification}.
\newblock \emph{\bibinfo{journal}{Phys. Rev. A}} \textbf{\bibinfo{volume}{75}},
  \bibinfo{pages}{052318} (\bibinfo{year}{2007}).

\bibitem{lloyd2001}
\bibinfo{author}{Lloyd, S.}, \bibinfo{author}{Shahriar, M.~S.},
  \bibinfo{author}{Shapiro, J.~H.} \& \bibinfo{author}{Hemmer, P.~R.}
\newblock \bibinfo{title}{Long distance, unconditional teleportation of atomic
  states via complete {B}ell state measurements}.
\newblock \emph{\bibinfo{journal}{Phys. Rev. Lett.}}
  \textbf{\bibinfo{volume}{87}}, \bibinfo{pages}{167903}
  (\bibinfo{year}{2001}).

\bibitem{bochmann2010}
\bibinfo{author}{Bochmann, J.} \emph{et~al.}
\newblock \bibinfo{title}{Lossless state detection of single neutral atoms}.
\newblock \emph{\bibinfo{journal}{Phys. Rev. Lett.}}
  \textbf{\bibinfo{volume}{104}}, \bibinfo{pages}{203601}
  (\bibinfo{year}{2010}).

\bibitem{volz2011}
\bibinfo{author}{Volz, J.}, \bibinfo{author}{Gehr, R.},
  \bibinfo{author}{Dubois, G.}, \bibinfo{author}{Estève, J.} \&
  \bibinfo{author}{Reichel, J.}
\newblock \bibinfo{title}{Measurement of the internal state of a single atom
  without energy exchange}.
\newblock \emph{\bibinfo{journal}{Nature}} \textbf{\bibinfo{volume}{475}},
  \bibinfo{pages}{210--213} (\bibinfo{year}{2011}).

\bibitem{duan2005}
\bibinfo{author}{Duan, L.-M.}, \bibinfo{author}{Wang, B.} \&
  \bibinfo{author}{Kimble, H.~J.}
\newblock \bibinfo{title}{Robust quantum gates on neutral atoms with
  cavity-assisted photon scattering}.
\newblock \emph{\bibinfo{journal}{Phys. Rev. A}} \textbf{\bibinfo{volume}{72}},
  \bibinfo{pages}{032333} (\bibinfo{year}{2005}).

\bibitem{nussmann2005}
\bibinfo{author}{Nußmann, S.} \emph{et~al.}
\newblock \bibinfo{title}{Submicron positioning of single atoms in a
  microcavity}.
\newblock \emph{\bibinfo{journal}{Phys. Rev. Lett.}}
  \textbf{\bibinfo{volume}{95}}, \bibinfo{pages}{173602}
  (\bibinfo{year}{2005}).

\end{thebibliography}
\end{document}